\documentclass[journal=nalefd,manuscript=letter]{achemso}

\usepackage{chemformula} 
\usepackage[T1]{fontenc} 
\usepackage{multirow}
\usepackage{subfigure} 
\usepackage{hyperref}
\usepackage{soul}
\usepackage{array}

\usepackage{color}

\author{Yanyu Duan}
\affiliation{Thrust of Advanced Materials, and Guangzhou Municipal Key Laboratory of Materials Informatics, The Hong Kong University of Science and Technology (Guangzhou), Guangzhou 511453, China.}

\author{Zecheng Gan} 
\email{zechenggan@ust.hk}
\affiliation{Thrust of Advanced Materials, and Guangzhou Municipal Key Laboratory of Materials Informatics, The Hong Kong University of Science and Technology (Guangzhou), Guangzhou 511453, China.}
\alsoaffiliation{Department of Mathematics, The Hong Kong University of Science and Technology, Hong Kong SAR, China}

\title[Coulombic Contact Interactions through Asymmetry]
  {Designing Coulombic Contact Interactions between Polarizable Particles through Asymmetry}

\keywords{dielectric spheres, polarization, image charges, self-assembly}

\begin{document}




\begin{abstract}
Polarizable particle systems, including charged colloids, polarizable ions, biomolecular assemblies, and soft nanomaterials, can exhibit contact electrostatic interactions that depart strongly from Coulomb behavior when dielectric mismatch and geometric singularities amplify polarization effects.
Here we use charged dielectric spheres as a model system and show that these polarization contributions can be canceled by jointly tuning size, charge, and dielectric asymmetries.
By extending a recently developed image-charge formula to contacting dielectric spheres, at the two-body level, we derive analytical conditions under which the two-body contact interaction reduces to the bare Coulomb form.
At the selected design points, accurate two-sphere calculations give root-mean-square (RMS) contact-force deviations below $3\%$ from the Coulomb target.
Separately, many-body molecular dynamics simulations that retain nonadditive polarization show that, under the tested assembly conditions, the structural correlations of six systems satisfying these two-body rules closely match those of their pure Coulomb references.
These results establish asymmetry as a route for turning electrostatic complexity into Coulombic simplicity at contact, with implications for predictable and controlled self-assembly and materials design.
\end{abstract}


\section{Introduction}

Electrostatic interactions are often treated as bare Coulomb interactions when charged particles approach one another, but this approximation can fail at contact in polarizable systems. 
Charged colloids, polarizable ions, biomolecular assemblies, and soft nanomaterials commonly contain dielectric interfaces, and the induced charges at these interfaces can reshape short-range forces that control aggregation, crystallization, and self-assembly~\cite{levin2002electrostatic, french2010long, walker2011electrostatics,SAleunissen2005ionic,SAkewalramani2016electrolyte}. A minimal model that isolates this mechanism consists of charged dielectric spheres immersed in a homogeneous medium whose permittivity differs from that of the particles~\cite{SAbarros2014dielectric}. 
Despite its simple geometry, this model captures the essential contact-scale challenge: dielectric mismatch and geometric singularities can amplify polarization contributions until the interaction departs strongly from the Coulomb limit.

These departures are not small corrections. Recent studies have shown that conductor-like particles, whose dielectric permittivity is higher than that of the surrounding medium, can exhibit short-range ``like-charge attraction'', especially under size or charge asymmetry~\cite{LCAbesley2023recent,LCAduan2025quantitative,Nigorman2024electrostatic}. Conversely, insulator-like particles can show ``opposite-charge repulsion'' under related conditions~\cite{LCAduan2025mechanisms}. 
Such effects make contact electrostatics difficult to predict and difficult to use as a design variable: the same nominal Coulomb interaction can be strengthened, weakened, or even effectively reversed by polarization. 
A central question is therefore whether polarization at contact must be treated only as a complication, or whether it can be deliberately canceled by tailoring size, charge, and dielectric asymmetries between the particles.

The electrostatic contact problem has a long history, beginning with Maxwell's analysis of conducting spheres and later work on weak singularities in conductor polarization energy~\cite{maxwell1873treatise,lekner2012electrostatics}. 
For dielectric particles, Qin and co-workers derived induced-charge equations in bispherical coordinates to describe the contact energy of identical dielectric spheres and later extended the analysis to clusters of contacting dielectrics~\cite{Bislian2018polarization,Bilian2022exact}. 
More broadly, substantial progress has been made in computing and understanding polarization in dielectric sphere systems, including bispherical-coordinate treatments of two-sphere contact~\cite{chan2015general}, polarizable ion models~\cite{Pochan2020theory}, multiple-scattering formalisms~\cite{MSFfreed2014perturbative,MSFqin2016theory,MSFqin2019charge}, boundary integral equations~\cite{barros2014efficient,gan2015comparison,lindgren2018integral}, image-charge methods~\cite{ICMxu2013electrostatic,ICMwang2013effects}, and the method of moments~\cite{MEbichoutskaia2010electrostatic}. 
However, these advances do not yet provide a simple materials design framework for contacting dielectric spheres with simultaneous size, charge, and dielectric asymmetries.

In this paper, we show that asymmetry can be used to program Coulombic contact interactions between polarizable particles. We first revisit a recently developed image-charge formula for polarizable spheres~\cite{LCAduan2025quantitative} and extend it to charged dielectric spheres in contact. 
The resulting analytical expression separates the bare Coulomb interaction from polarization contributions controlled by size, charge, and dielectric asymmetries, allowing us to derive cancellation conditions under which the contact energy reduces to the Coulomb form. 
We validate these contact design rules against accurate Hybrid Method simulations~\cite{HMgan2016hybrid,HMgan2019efficient}; for every set of four validation points with a fixed charge or size asymmetry, the RMS contact-force deviation from the pure Coulomb value is below $3\%$. For six designed parameter sets, many-body molecular dynamics simulations further show that the radial distribution functions closely follow their pure Coulomb references under assembly conditions.

The broader significance of this inverse design formulation lies in materials selection for self-assembly. Once the particle materials and suspending medium are chosen, their dielectric permittivities are largely fixed, and dielectric mismatch may change the intended contact interaction. Particle size and net charge, however, can often be varied through synthesis and surface functionalization. For a chosen particle--medium combination, the analytical conditions identify size or charge asymmetries that compensate the leading contact polarization correction. This creates the possibility of selecting particles for desired optical, chemical, or mechanical functionality while retaining Coulomb-like contact interactions as a basis for assembly design. More broadly, the framework may extend self-assembly strategies developed for ideal Coulombic particles to a wider range of polarizable materials. The present results establish this possibility for the tested systems under assembly conditions; applying it to more specific target structures will require dedicated experimental and computational validation.

\section{Theory}

The design objective is to make the electrostatic contact energy between two polarizable spheres recover the bare Coulomb energy,
$E_{\mathrm{ele}}=E_{\mathrm{Coul}}$, even when the spheres have dielectric mismatch with the surrounding medium.
To reach this objective analytically, the theory must separate the contact energy into a direct Coulomb term and a polarization correction whose sign and magnitude can be tuned by material and geometric asymmetry.
We obtain this separation by extending a recently developed compact image-charge formula for dielectric spheres~\cite{LCAduan2025quantitative} to the contact geometry.

\begin{figure}[htbp]
    \centering
    \subfigure[]{
        \includegraphics[width=0.4\textwidth]{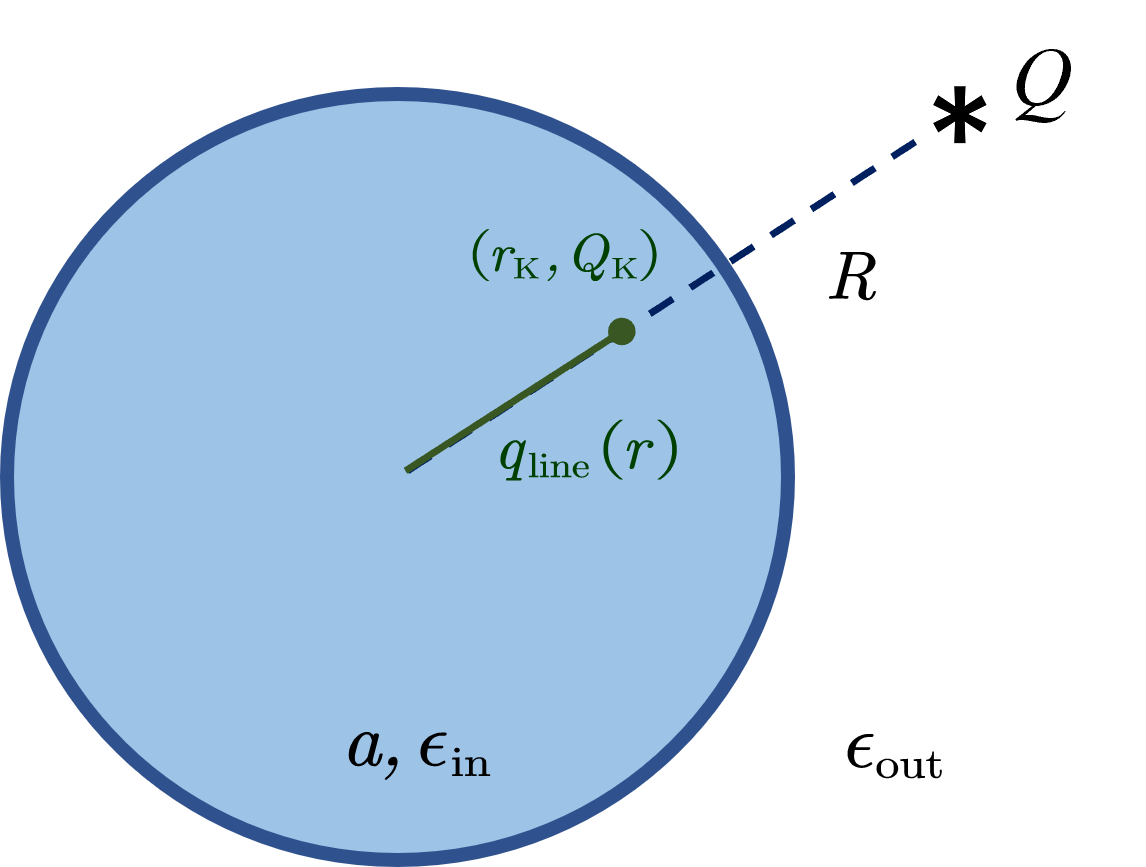}
        \label{schematic}
    }
    \subfigure[]{
        \includegraphics[width=0.4\textwidth]{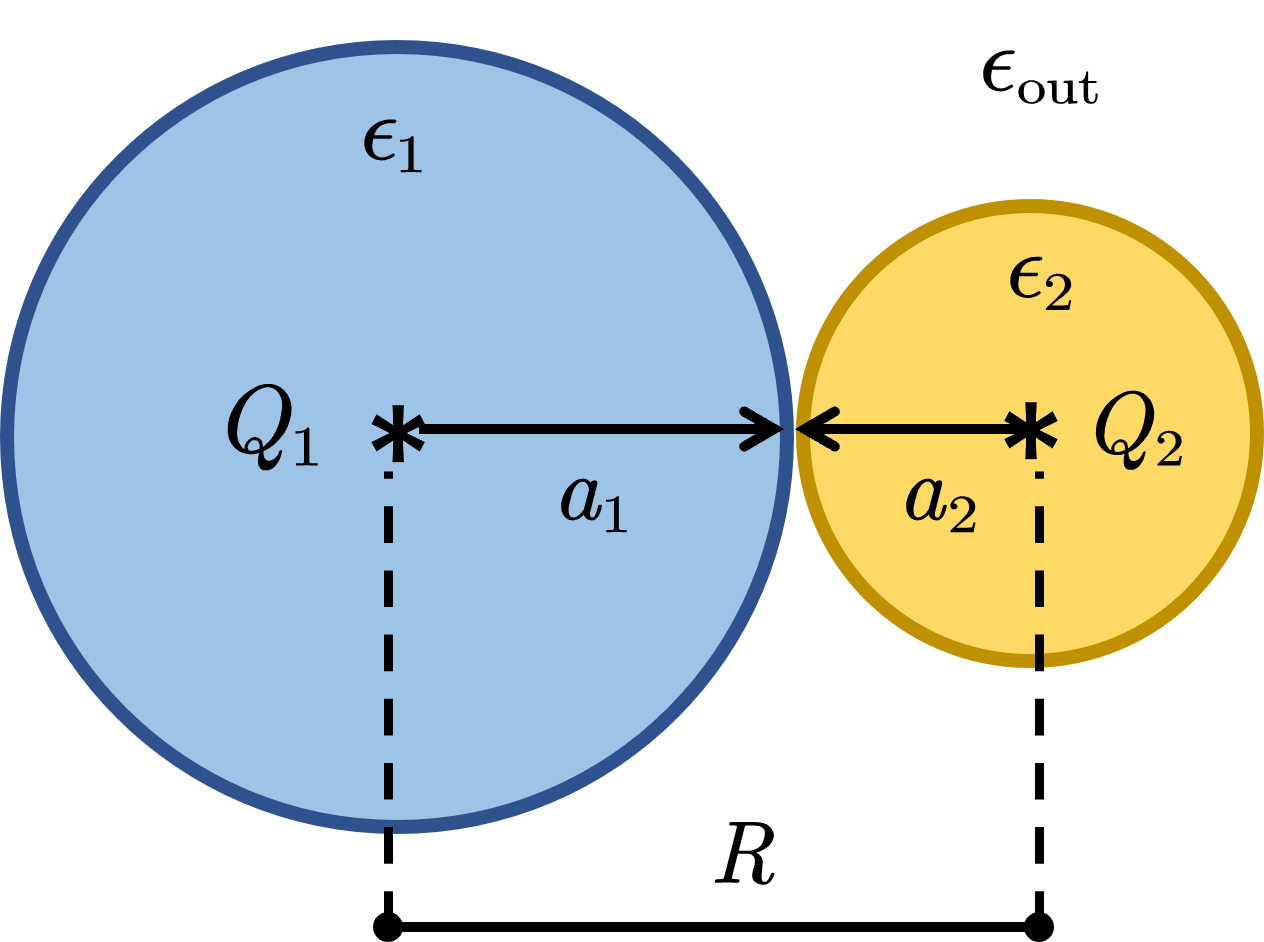}
        \label{sche}
    }
    \caption{Schematics of the dielectric sphere model. (a) A point charge $Q$ interacts with a neutral dielectric sphere of radius $a$ and dielectric constant $\epsilon_{\text{in}}$ at separation $R$. According to Neumann's image principle, the polarization potential is generated by a Kelvin image charge $Q_{\text{K}}$ at $r_{\text{K}}$ and a line image charge density $q_\text{line}(r)$ distributed along the green line connecting the Kelvin point and the sphere center. (b) A pair of charged dielectric spheres in contact, with radii ($a_1$, $a_2$), dielectric constants ($\epsilon_1$, $\epsilon_2$), and central charges ($Q_1$, $Q_2$). Both systems are immersed in a medium with permittivity $\epsilon_{\text{out}}$.}
    \label{fig:main1}
\end{figure}

We first recall the single-sphere building block. Consider a point charge $Q$ interacting with a neutral dielectric sphere of radius $a$ and dielectric constant $\epsilon_{\text{in}}$, both immersed in a homogeneous medium with permittivity $\epsilon_{\text{out}}$, as depicted in Fig.\ref{schematic}. 
According to the classical \emph{Neumann image principle}~\cite{neumann1883hydrodynamische,lindell1993application}, the polarization potential outside the sphere can be represented by a Kelvin image charge $Q_{\text{K}}=-\frac{Qka}{R}$ located at $r_{\text{K}}=\frac{a^2}{R}$ and a Neumann image line with charge density $q_\text{line}(r)=\frac{Qkg}{a}\left(\frac{r_\text{K}}{r}\right)^{1-g}$ distributed from the sphere center to the Kelvin point.
Here the dimensionless dielectric contrast is $k=\frac{\epsilon_{\text{in}}-\epsilon_{\text{out}}}{\epsilon_{\text{in}}+\epsilon_{\text{out}}}$, and $g=\frac{1}{1+\frac{\epsilon_{\text{in}}}{\epsilon_{\text{out}}}}$.
The polarization energy $E_{\text{pol}}$ for the single-sphere system is then
\begin{equation}
    E_{\mathrm{pol}} =\frac{1}{2}\frac{Q}{4\pi\varepsilon_0\epsilon_\text{out}} \left[ \frac{Q_\text{K}}{R-r_\text{K}} + \int_{0}^{r_\text{K}} \frac{q_\text{line}(r)}{R-r} \, dr\right]\;,
    \label{ICM1}
\end{equation}
where $\varepsilon_0$ is the vacuum permittivity, and the extra prefactor $1/2$ accounts for the fictitious nature of image charges.
In the perfect-conductor limit, i.e., $\epsilon_{\text{in}}\to +\infty$, the line charge density degenerates to a point charge at the sphere center and Eq.~\eqref{ICM1} reduces to the classical image formula for a spherical conductor~\cite{jackson2021classical}.
For a finite dielectric constant $\epsilon_{\text{in}}$, however, direct use of the image-line integral is inconvenient because the line charge is singular at the sphere center ($r=0$). Although tailored discretization schemes enable this representation in Monte Carlo (MC) and molecular dynamics (MD) simulations~\cite{cai2007extending,gan2011multiple,gan2015comparison}, the integral form does not directly lead to a simple analytical design rule for canceling the contact polarization correction.

By defining the dimensionless separation variable $t=\frac{a}{R}$, Eq.~\eqref{ICM1} can be equivalently written as~\cite{LCAduan2025quantitative}
\begin{equation}
    E_{\mathrm{pol}} = \frac{1}{2} \frac{Q}{4\pi\varepsilon_0\epsilon_\text{out}R} \left[ -\frac{Qkt}{1-t^2} + \frac{Qktg}{t^{2g}} B_{t^2}(g,0)\right]\;,
    \label{ICM2}
\end{equation}
where $B_{t^2}(g,0)$ is the incomplete beta function~\cite{arfken2011mathematical}.
Using the series expansion for $B_{t^2}(g,0)$~\cite{pearson1968tables},
\begin{equation}
	\begin{split}
		B_{t^2}(g,0)&=t^{2g} \sum_{n=0}^{\infty}\frac{t^{2n}}{n+g}\;,
	\end{split}
	\label{beta}
\end{equation}
and substituting $g=\frac{1-k}{2}$, Eq.~\eqref{ICM2} can be simplified as
\begin{equation}
    E_{\mathrm{pol}} = -\frac{1}{2} \frac{Q^2kt}{4\pi\varepsilon_0\epsilon_\text{out}R} \left[ \frac{1}{1-t^2} - (1-k)\sum_{n=0}^{\infty}\frac{t^{2n}}{2n+1-k}\right]\;.
    \label{ICM3}
\end{equation}
Equation~\eqref{ICM3} is mathematically equivalent to Neumann's image principle, but it replaces the singular image-line integral in Eq.~\eqref{ICM1} with a simple series. 
This series form is the key step that makes the contact-cancellation problem analytically tractable.

We now consider two dielectric spheres in contact, as shown in Fig.\ref{sche}. The spheres may have different radii ($a_1$, $a_2$), dielectric constants ($\epsilon_1$, $\epsilon_2$), and central charges ($Q_1$, $Q_2$), with center-to-center distance $R=a_1+a_2$.
Applying Eq.~\eqref{ICM3} to each sphere, and treating the central charge in the other sphere as the external source charge, gives the first-reflection contact energy
\begin{equation}
		\begin{split}
           E_{\mathrm{ele}} =\frac{Q_1Q_2}{4\pi\varepsilon_0\epsilon_\text{out}
    (a_1+a_2)} \left\{1  -  \frac{k_2t_2}{2}\frac{Q_1}{Q_2}
     \left[ \frac{1}{1-t_2^2} - (1-k_2) \sum_{n=0}^{\infty}\frac{t_2^{2n}}{2n+1-k_2}\right] 
     \right.  \\ \left.      
    -\frac{k_1t_1}{2}\frac{Q_2}{Q_1} \left[\frac{1}{1-t_1^2} - (1-k_1)\sum_{n=0}^{\infty}\frac{t_1^{2n}}{2n+1-k_1}\right] \right\}
		\end{split}\;,
		\label{energy}
\end{equation}
where $k_1=\frac{\epsilon_1-\epsilon_{\text{out}}}{\epsilon_1+\epsilon_{\text{out}}}$, $k_2=\frac{\epsilon_2-\epsilon_{\text{out}}}{\epsilon_2+\epsilon_{\text{out}}}$, $t_1=\frac{a_1}{a_1+a_2}$, $t_2=\frac{a_2}{a_1+a_2}$, and clearly $t_1+t_2\equiv 1$. 
Equation~\eqref{energy} provides the desired decomposition: the prefactor is the bare Coulomb interaction,
$E_{\mathrm{Coul}}=\frac{Q_1Q_2}{4\pi\varepsilon_0\epsilon_\text{out}(a_1+a_2)}$, and the bracket contains one plus a polarization correction.
Equivalently, $E_{\mathrm{ele}}/E_{\mathrm{Coul}}=1+\Phi$, where $\Phi$ is controlled only by the dielectric contrasts $k_{1,2}$, the contact size ratios $t_{1,2}$, and the charge ratio $Q_1/Q_2$.
The contact-design condition is therefore simply $\Phi=0$: the polarization terms generated by the two spheres must cancel each other.
This formulation turns size, charge, and dielectric asymmetries into explicit control parameters rather than uncontrolled sources of deviation from Coulomb behavior.

Although the model in Fig.~\ref{sche} represents each particle charge by a central point charge, Eq.~\eqref{energy} is unchanged if that charge is replaced by a fixed, uniform surface charge density $\sigma_i=Q_i/(4\pi a_i^2)$. The uniform surface charge produces the same exterior monopole potential, and its coupling to any source-free external potential equals $Q_i$ times the potential at the sphere center. The particle permittivity remains relevant because the nonuniform field generated by the other particle polarizes the dielectric interface through $k_i$. This equivalence does not apply to nonuniform, patchy, mobile, or charge-regulating surface charges, which introduce additional multipoles or charge redistribution and require a separate treatment.

Equation~\eqref{energy} is a leading-order expression for the contact energy because it truncates the image-reflection series at the first reflection.
Higher-order reflections give a more complete two-sphere polarization energy and have been used in numerical simulation studies~\cite{ICMxu2013electrostatic,li2024like,li2025electrostatic}.
Here the purpose of Eq.~\eqref{energy} is not to serve as a high-precision electrostatic solver, but to provide an analytical route to the cancellation conditions used for materials design.
To assess the effect of higher-order image reflections at contact, we compare the leading-order expression with the high-accuracy contact-limit values reported by Lian and Qin~\cite{Bislian2018polarization}. Their bispherical-coordinate calculation evaluates the two-sphere interaction energy at small positive separations and extrapolates it to contact, thereby retaining the full two-sphere polarization response. For each dielectric contrast in Table S1 of the \emph{Supporting Information} (SI), we define the relative contact-energy error as $\delta_E=|E_{\mathrm{ele}}-E_{\mathrm{Ref}}|/|E_{\mathrm{Ref}}|\times100\%$, where $E_{\mathrm{ele}}$ is given by Eq.~\eqref{energy} and $E_{\mathrm{Ref}}$ is the benchmark contact energy. All values of $\delta_E$ are below $5\%$. This comparison quantifies the energy-level error caused by the leading-order truncation.

The contact-energy error does not directly measure how closely the predicted cancellation points recover the Coulomb force. We therefore evaluate these points using the full Hybrid Method. For each set of four points with a fixed charge or size asymmetry, Tables S2 and S3 report the RMS contact-force deviation $\delta_F=\sqrt{\frac{1}{4}\sum_{i=1}^{4}(F_{\mathrm{Simul}}^{i}-F_{\mathrm{Coul}})^2}/|F_{\mathrm{Coul}}|\times100\%$, where $F_{\mathrm{Simul}}^{i}$ is the Hybrid Method force and $F_{\mathrm{Coul}}$ is the target Coulomb force. All tested sets have $\delta_F<3\%$.

\section{Role of dielectric asymmetry}
To expose the cancellation mechanism in its simplest form, we first remove charge and size asymmetry and isolate dielectric asymmetry alone; the isolated effects of charge and size asymmetry are discussed in the SI.
Here we set $Q_1=Q_2=Q$ and $a_1=a_2=a$, so that $t_1=t_2=\frac{1}{2}$.
In this case, the contact energy Eq.~\eqref{energy} becomes
\begin{equation}
\begin{split}
    E_{\mathrm{ele}} =
    \frac{Q^2}{4\pi\varepsilon_0\epsilon_\text{out}(2a)}
    \left[
    1-\frac{k_1+k_2}{3}
    +\frac{k_1(1-k_1)}{4}\sum_{n=0}^{\infty}
    \frac{4^{-n}}{2n+1-k_1}\right. \\
    \left.
    +\frac{k_2(1-k_2)}{4}\sum_{n=0}^{\infty}
    \frac{4^{-n}}{2n+1-k_2}
    \right]\;.
\end{split}
\label{permittivity}
\end{equation}
Equivalently, the normalized contact energy can be written as
\begin{equation}
    \frac{E_{\mathrm{ele}}}{E_{\mathrm{Coul}}}
    =1+\psi(k_1)+\psi(k_2)\;,
    \qquad
    \psi(k)=
    -\frac{k}{3}
    +\frac{k(1-k)}{4}\sum_{n=0}^{\infty}
    \frac{4^{-n}}{2n+1-k}\;.
    \label{dielectric_decomposition}
\end{equation}
This decomposition makes the role of dielectric asymmetry transparent.
The function $\psi(k)$ is zero at $k=0$, positive for insulator-like particles ($-1< k<0$), and negative for conductor-like particles ($0<k<1$).
Thus two particles with dielectric contrasts of the same sign push the contact energy to the same side of the Coulomb value and cannot cancel each other's polarization contributions.
Cancellation requires the two contributions in Eq.~\eqref{dielectric_decomposition} to have opposite signs, which means that one sphere must be conductor-like and the other insulator-like.

\begin{figure}[htbp]
    \centering
    \includegraphics[scale=1]{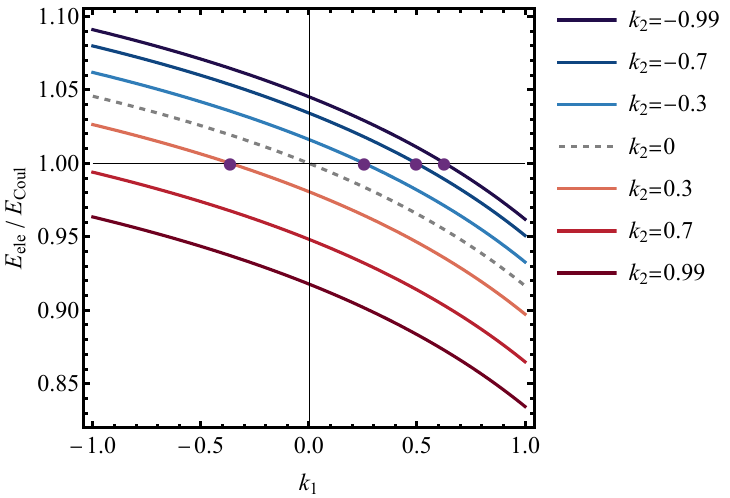}
    \caption{Influence of dielectric asymmetry ($k_1\neq k_2$) on the contact energy. The normalized contact energy ($E_{\mathrm{ele}}/E_{\mathrm{Coul}}$) is plotted versus $k_1$ for different $k_2$. Filled circles indicate nontrivial $(k_1, k_2)$ pairs for which the polarization correction vanishes.}
    \label{dielectric}
\end{figure}

In Fig.~\ref{dielectric}, we plot the contact energy (normalized by the pure Coulomb energy $E_{\mathrm{Coul}}$) versus $k_1$, with varying $k_2\in(-1, 1)$.
For fixed $k_2$, the contact energy decreases monotonically as $k_1$ (or $\epsilon_1$) increases, consistent with the monotonic decrease of $\psi(k_1)$.
When $k_1$ and $k_2$ have the same sign ($k_1k_2>0$), the contact energy is weakened relative to the Coulomb energy for two conductor-like spheres ($k_1>0$, $k_2>0$) and enhanced for two insulator-like spheres ($k_1<0$, $k_2<0$).
When $k_1k_2<0$, the two polarization corrections compete, so the contact energy can lie above, below, or exactly at the Coulomb limit depending on their relative magnitudes.
The filled circles in Fig.~\ref{dielectric} mark the nontrivial solutions of $\psi(k_1)+\psi(k_2)=0$, where the contact energy returns to $E_{\mathrm{Coul}}$ even though both spheres remain polarizable.
Dielectric asymmetry alone therefore already reveals the basic design principle: the contact energy can recover the bare Coulomb value even with finite dielectric mismatch, provided that the two spheres generate polarization corrections of opposite sign.
Because $k_i$ is defined relative to the solvent, the requirement $k_1k_2<0$ means that the two particle permittivities lie on opposite sides of $\epsilon_{\mathrm{out}}$. This requirement is more demanding in water, whose large static dielectric constant gives $k<0$ for many common colloidal materials. High-permittivity ferroelectric oxides such as BaTiO$_3$ may provide positive contrast, although their effective nanoparticle permittivity depends on particle size and interfacial structure and should be characterized for the system of interest~\cite{padurariu2023microstructures}. Lower-permittivity nonaqueous media provide another route and have been used in charged-colloid model systems~\cite{SAleunissen2005ionic}. Applying the design rule therefore requires an experimentally accessible positive-contrast particle.
For equal charges and equal sizes, however, this condition defines only a narrow cancellation curve on the $(k_1,k_2)$ plane.
We next show how further introducing charge or size asymmetry expands this curve into a broader design space.

\section{Design rules for Coulombic contact interactions}
The cancellation condition identified above can be generalized by returning to Eq.~\eqref{energy} and allowing additional asymmetry to tune the two polarization terms.
We now derive explicit design rules for recovering $E_{\mathrm{ele}}=E_{\mathrm{Coul}}$ by simultaneously adjusting dielectric asymmetry with either charge or size asymmetry.
For clarity, we consider two experimentally transparent scenarios: (a) two equal-sized spheres with both charge and dielectric asymmetries; and (b) two unequal-sized spheres with dielectric asymmetry and equal charge magnitudes.

\begin{figure}[htbp]
    \centering
    \subfigure[]{
        \includegraphics[width=0.54\textwidth]{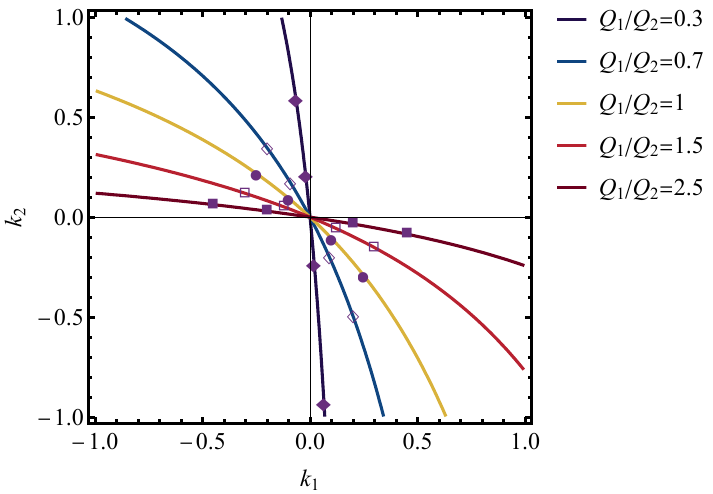}
        \label{zero}
    }
    \subfigure[]{
        \includegraphics[width=0.4\textwidth]{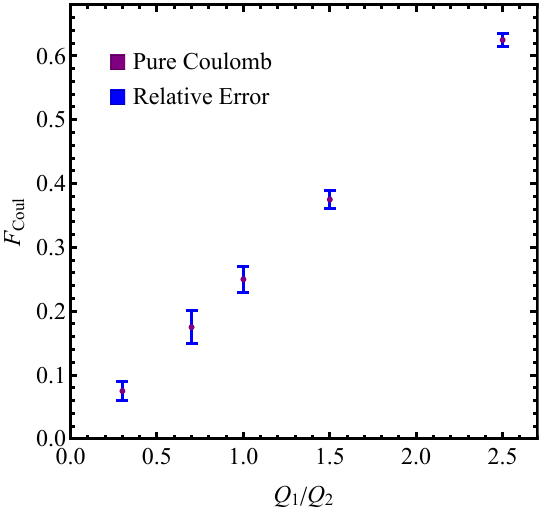}
        \label{coul}
    }
    \caption{
    (a) Predicted design rules for recovering the pure Coulomb contact interaction by jointly tuning charge and dielectric asymmetries. 
    Feasible sets of dielectric contrast pairs $(k_1, k_2)$ are plotted for different charge asymmetries $Q_1/Q_2$.
    Markers indicate selected validation data points. (b) Numerical validation by comparing the simulated electrostatic contact force for the designed parameters with the pure Coulomb force $F_{\mathrm{Coul}}$; error bars are calculated from the validation data marked in Fig.~\ref{zero}, with details listed in Table S2 of the SI.
    }
    \label{fig:main3}
\end{figure}

The design rule is obtained by setting the polarization correction in Eq.~\eqref{energy} to zero.
To keep the cancellation conditions transparent, we define the single-sphere contact polarization response
\[
    \mathcal{P}(k,t)=
    kt\left[
    \frac{1}{1-t^2}
    -(1-k)\sum_{n=0}^{\infty}
    \frac{t^{2n}}{2n+1-k}
    \right]\;.
\]
This response depends on both the dielectric contrast $k$ and the contact size ratio $t$; its sign distinguishes conductor-like and insulator-like particles.
For scenario (a), $t_1=t_2=1/2$, and charge asymmetry changes only the relative weights of the two dielectric polarization terms.
The charge ratio $Q_1/Q_2$ and dielectric contrasts $k_1$ and $k_2$ must therefore satisfy
\begin{equation}
    \left(\frac{Q_1}{Q_2}\right)^2
    \mathcal{P}\left(k_2,\frac{1}{2}\right)
    +\mathcal{P}\left(k_1,\frac{1}{2}\right)=0\;.
\label{condition}
\end{equation}
For a fixed charge ratio, Eq.~\eqref{condition} gives the dielectric-contrast pairs $(k_1,k_2)$ for which the two weighted polarization corrections cancel.
Changing $|Q_1/Q_2|$ therefore shifts the cancellation curve in the $(k_1,k_2)$ plane.
Because only $\left(Q_1/Q_2\right)^2$ appears, the contact-energy design rule is independent of whether the two particles are like charged or oppositely charged.
We therefore plot the feasible sets in Fig.~\ref{zero} using $Q_1/Q_2>0$ without loss of generality.
The series defining $\mathcal{P}$ is rapidly convergent; truncating at $n_{\mathrm{max}}=10$ is sufficient for all curves shown here.

\begin{figure}[htbp]
    \centering
    \subfigure[]{
        \includegraphics[width=0.53\textwidth]{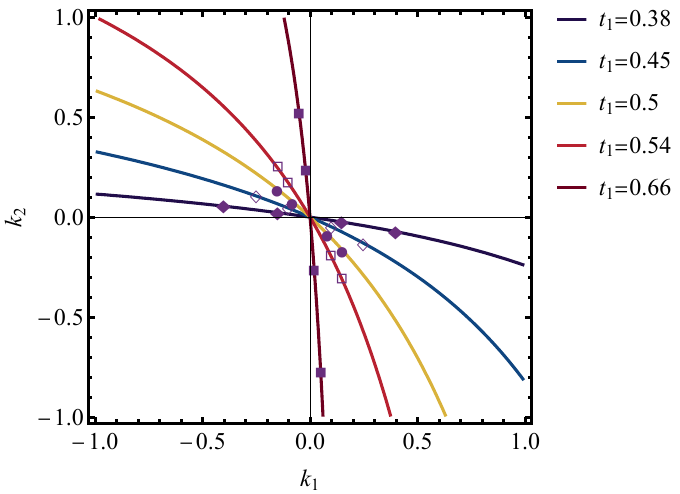}
        \label{unequal}
    }
    \subfigure[]{
        \includegraphics[width=0.42\textwidth]{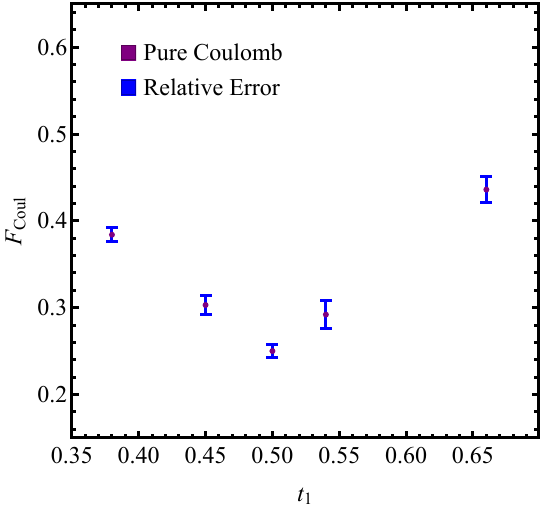}
        \label{compare}
    }
    \caption{(a) Predicted design rules for recovering the pure Coulomb contact interaction by jointly tuning size and dielectric asymmetries. 
    Feasible sets of dielectric contrast pairs $(k_1, k_2)$ are plotted for different size asymmetries $t_1$.
    Markers indicate selected validation data points. (b) Numerical validation by comparing the simulated electrostatic contact force for the designed parameters with the pure Coulomb force $F_{\mathrm{Coul}}$; error bars are calculated from the validation data marked in Fig.~\ref{unequal}, with details listed in Table S3 of the SI.}
    \label{fig:main4}
\end{figure}

For scenario (b), we set equal charge magnitudes and use size asymmetry to tune the two polarization responses.
Sphere 1 contributes through the contact ratio $t_1$, whereas sphere 2 contributes through $t_2=1-t_1$.
The cancellation condition is therefore
\begin{equation}
    \mathcal{P}(k_2,1-t_1)+\mathcal{P}(k_1,t_1)=0\;.
\label{condition1}
\end{equation}
For a fixed $t_1$, Eq.~\eqref{condition1} gives the dielectric-contrast pairs $(k_1,k_2)$ that recover the Coulomb contact interaction.
Changing $t_1$ changes the relative weights of the two single-sphere responses, so each size ratio produces a different cancellation curve, as shown in Fig.~\ref{unequal}.

The cancellation curves identify exact parameter combinations. If the dielectric contrasts, charge ratio, or size ratio deviate from a selected point, the two polarization contributions no longer cancel completely and a residual correction remains. Polydispersity therefore weakens the exact cancellation because particle pairs with different radii have different values of $t_1=a_1/(a_1+a_2)$ and cannot all satisfy the same cancellation condition. The present simulations use monodisperse particles and therefore do not determine how much polydispersity can be tolerated. For a material system with known size and charge distributions, Eq.~\eqref{energy} can be evaluated for particle pairs sampled from those distributions to estimate the range of residual contact corrections. Many-body simulations that explicitly include the same distributions would then be needed to determine whether these pair-level deviations affect the assembled structure. When the available dielectric constants are fixed, the mean size or charge ratio may still be chosen to bring the system closer to a cancellation curve, although this does not eliminate the effect of polydispersity.

Figures~\ref{zero} and~\ref{unequal} show that, in both scenarios, all feasible dielectric-contrast pairs $(k_1,k_2)$ lie in the second and fourth quadrants.
This confirms the necessary condition anticipated from Eq.~\eqref{dielectric_decomposition}: canceling the contact polarization correction requires $k_1k_2<0$, so one sphere must be conductor-like and the other insulator-like.
When only dielectric asymmetry is present ($Q_1/Q_2=1$ in Fig.~\ref{zero} and $t_1=0.5$ in Fig.~\ref{unequal}), the feasible set reduces to the narrow cancellation curve identified in Fig.~\ref{dielectric}.
Introducing charge or size asymmetry reweights the two competing polarization corrections and expands this curve into a broader family of dielectric design rules while preserving the sign requirement $k_1k_2<0$.

To validate the design rules predicted by Eqs.~\eqref{condition} and~\eqref{condition1}, we selected representative parameter sets labeled in Figs.~\ref{zero} and~\ref{unequal} and computed the corresponding interaction forces with an accurate numerical method~\cite{HMgan2019efficient}. 
Because Eqs.~\eqref{condition} and~\eqref{condition1} are derived from the contact energy, we further test whether the same designed parameters also reproduce the Coulomb force at contact.
The computed electrostatic forces are compared with the pure Coulomb force $F_{\mathrm{Coul}}$ in Figs.~\ref{coul} and~\ref{compare} for scenarios (a) and (b), respectively; each error bar is obtained from four parameter sets with fixed charge or size asymmetry.
For both scenarios, the RMS contact-force deviations $\delta_F$ are below $3\%$ (Tables S2 and S3 of the SI), showing that the selected design points remain close to the Coulomb-force target when the full polarization response is included.

To give a concrete materials-parameter illustration, consider metallic gold and fused-silica spheres in chloroform at $20~^{\circ}\mathrm{C}$. Taking $\epsilon_{\mathrm{out}}=4.81$ for chloroform and $\epsilon_2\approx3.90$ for fused silica, together with the conductor limit $\epsilon_1\to\infty$, gives $k_1=1$ and $k_2\approx-0.10448$.~\cite{lide2005crc,young1973dielectric} For equal fixed charge magnitudes, Eq.~\eqref{condition1} gives $t_1=0.33132$, or $a_2/a_1\approx2.018$. Thus, a gold sphere with $a_1=50~\mathrm{nm}$ is paired with a fused-silica sphere of $a_2\approx100.9~\mathrm{nm}$. This size asymmetry is within reach of standard synthesis.

\clearpage
\section{Self-assembly with designed Coulomb interactions}

Matching the Coulomb contact interaction is neither a necessary nor a sufficient condition for self-assembly, which depends on the balance of energetic and entropic contributions to the free energy.
Here, the pure Coulomb systems serve only as structural references.
The pair-derived condition $\Phi=0$ removes the leading contact-scale polarization correction for an isolated pair but leaves higher-order two-sphere reflections and many-particle polarization.
We therefore use many-body MD with the Hybrid Method~\cite{HMgan2019efficient} to test whether the designed polarizable systems reproduce the reference structural correlations under the assembly conditions examined here.

Briefly, the Hybrid Method represents the induced surface polarization with the method of moments (using spherical harmonics), resolves near-contact polarization with image charges, and accelerates the remaining long-range interactions using the fast multipole method~\cite{HMgan2016hybrid,HMgan2019efficient}. At each MD step, the induced polarization is solved self-consistently for the instantaneous particle configuration, and the resulting direct and induced electrostatic forces are used to propagate the particles.

Each system contains $200$ charged dielectric spheres at a total volume fraction of $3\%$.
The temperature is controlled by a Langevin thermostat, and the coupling strength is fixed at $\lambda = |E_{\mathrm{Coul}}|/(k_\text{B}T)=100$, where $E_{\mathrm{Coul}} = \frac{Q_1 Q_2}{4\pi\varepsilon_0 \epsilon_\text{out} (a_1 + a_2)}$ is the direct Coulomb interaction between two spheres in contact.

The simulations used an in-house implementation of the Hybrid Method and a spherical simulation cell whose radius was chosen to give the stated total particle volume fraction of $3\%$. Following established simulations of charged dielectric colloids~\cite{SAbarros2014dielectric,HMgan2019efficient}, particle overlap was prevented by a purely repulsive shifted-truncated Lennard--Jones potential with energy scale $\epsilon_{\mathrm{LJ}}=k_{\mathrm{B}}T$ and pair-dependent shift $\Delta_{ij}=a_i+a_j-\sigma$. All particles were assigned a mass $m_0$. We define the time unit as $t_0=\sigma\sqrt{m_0/(k_{\mathrm{B}}T)}$, where $\sigma$ is the smaller-particle diameter for unequal-sized systems and the common diameter for equal-sized systems.

For the electrostatic calculations, we used spherical-harmonic truncation order $p=5$, image cutoff $\eta=6$, nine-digit FMM accuracy, a GMRES tolerance of $10^{-6}$, and three-point Gauss--Jacobi quadrature for the image-line integrals~\cite{HMgan2019efficient}. We propagated the particle coordinates with the standard velocity-Verlet algorithm using a timestep $\Delta t=0.001t_0$ and controlled the temperature with a Langevin thermostat with damping time $t_0$. Each trajectory started from a random nonoverlapping particle configuration, with velocities drawn randomly from the Maxwell--Boltzmann distribution at the target temperature. Each simulation ran for $10^6$ steps. We discarded the first $2\times10^5$ steps, which cover the initial relaxation of the total electrostatic energy shown in Fig. S3 of the SI, and used the remaining $8\times10^5$ steps for production. We sampled the energies and RDFs every $500$ steps. For each parameter set, we performed five independent trajectories with independently generated initial configurations and velocities. The RDF curves in Fig.~\ref{fig3} report the mean over these five trajectories, and the error bars for all systems represent one standard deviation across the trajectories. The snapshots provide qualitative illustrations only.
The simulation source code, complete setup files for each simulated parameter set, and the trajectory-resolved RDF data underlying Fig.~\ref{fig3} are publicly available at \emph{https://github.com/zcgan/DesignRule-Code-Data}.

For the equal-sized systems, each species contains 100 spheres, with $Q_1/Q_2=-1$ and $a_1/a_2=1$.
The pure Coulomb reference (i) uses $(\epsilon_1,\epsilon_2)=(1,1)$, while the designed systems use (ii) $(1.35294,0.71782)$, (iii) $(0.54761,1.66667)$, and (iv) $(0.63106,1.50000)$, giving three distinct points on the dielectric-asymmetry cancellation curve defined by Eq.~\eqref{condition}.
For the unequal-sized systems, each species again contains 100 spheres with $Q_1/Q_2=-1$, while $a_1/a_2=0.613$.
The pure Coulomb reference (i) uses $(\epsilon_1,\epsilon_2)=(1,1)$, while the designed systems use (ii) $(2.33333,0.86446)$, (iii) $(0.73913,1.04554)$, and (iv) $(1.50000,0.93499)$, giving three distinct points on the size--dielectric cancellation curve defined by Eq.~\eqref{condition1}.

\begin{figure}[htbp]
    \centering
    \subfigure[]{
        \includegraphics[width=0.95\textwidth]{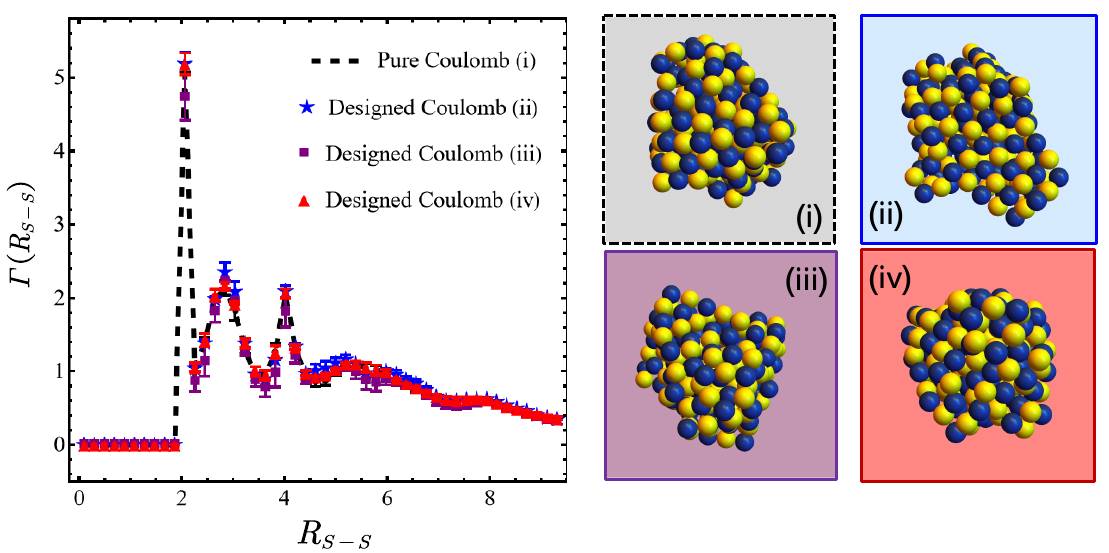}
        \label{111}
    }
    \subfigure[]{
        \includegraphics[width=0.95\textwidth]{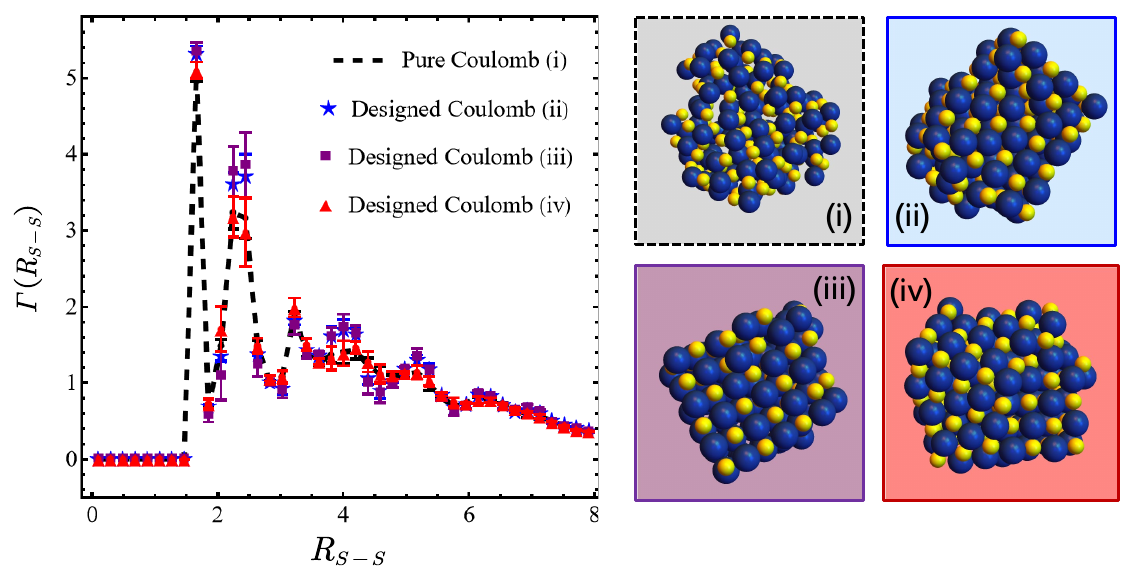}
        \label{2222}
    }
    \caption{Many-body test of designed Coulombic contact interactions. Radial distribution functions for (a) equal-sized and (b) unequal-sized oppositely charged spheres ($N=200$) compare a pure Coulomb reference (i, dashed lines) with three distinct designed polarizable systems (ii--iv, symbols) selected at separated points along the relevant cancellation curves [Eqs.~\eqref{condition} and~\eqref{condition1}].
    Under assembly conditions (a total volume fraction of $3\%$ and strong coupling $\lambda=100$), the designed systems closely follow the structural correlations of the pure Coulomb references, while the representative snapshots exhibit qualitatively similar aggregate morphologies.
    The RDF curves show the mean over five independent trajectories, and the error bars for all systems represent one standard deviation across these trajectories.} $R_{s-s}$: center-to-center distance between spheres with opposite charges.
    \label{fig3}
\end{figure}

The radial distribution functions (RDFs) $\Gamma(R_{s-s})$ for the equal-sized and unequal-sized systems are shown in Figs.~\ref{111} and~\ref{2222}, respectively.
Under these assembly conditions, the RDFs of all six designed parameter sets show similar structural correlations to their pure Coulomb references, while the representative configurations exhibit qualitatively similar aggregate morphologies.
To quantify this agreement, we computed the relative root-mean-square error (relative RMSE) between the RDF of each designed system and that of its corresponding pure Coulomb reference. For the equal-sized designed systems (ii--iv), the relative RMSEs are $6.415\%$, $9.337\%$, and $3.275\%$, respectively; for the unequal-sized designed systems (ii--iv), they are $10.649\%$, $13.935\%$, and $3.964\%$, respectively.
Note that, if collective polarization or simultaneous multiparticle contacts generate appreciable three-body or higher-order contributions, the pair-derived rule may lose accuracy and require an explicit many-body treatment.~\cite{nitzke2024long,nitzke2026quantitative,lindgren2025charge}.

\clearpage
\section{Conclusion}

This work establishes asymmetry as a design variable for recovering Coulombic contact interactions between polarizable particles.
By extending a compact image-charge formula to contacting dielectric spheres, we separated the bare Coulomb interaction from the leading polarization correction controlled by size, charge, and dielectric asymmetries.
Within the two-body theory, this decomposition yields analytical cancellation conditions under which the contact energy returns to the Coulomb value even though both particles remain dielectric mismatched with the surrounding medium.

The resulting design rules show that cancellation requires opposite-sign dielectric contrasts, $k_1k_2<0$, so that one sphere contributes a conductor-like correction and the other an insulator-like correction.
Charge or size asymmetry then reweights these competing contributions and expands the narrow dielectric-asymmetry cancellation curve into a broader family of feasible material parameters.
At the selected design points, accurate two-sphere calculations give RMS contact-force deviations below $3\%$ from the pure Coulomb values.
Separately, under the assembly conditions examined here, the many-body simulations for all six designed parameter sets yield RDFs that closely follow their pure Coulomb references, while representative configurations exhibit qualitatively similar aggregate morphologies.

The present framework assumes fixed particle charges in an ion-free, homogeneous dielectric medium. The equal numbers of oppositely charged particles make the simulated assemblies globally electroneutral, but this charge balance does not reproduce screening by mobile ions. In an electrolyte, mobile ions screen the field of the particle charges and redistribute near dielectric interfaces~\cite{Naturelee2023direct,YUANyuan2021particle,becker2025dielectric}, thereby changing both the direct interaction and the induced polarization. The present cancellation conditions therefore cannot be applied unchanged. In the weak-coupling, weak-screening limit, where the linearized Poisson--Boltzmann description is valid and $\kappa(a_1+a_2)\ll1$, the present result is recovered as the unscreened limit.

Within linear response, an electrolyte extension could combine the linearized Poisson--Boltzmann equation with screened image-charge or numerical methods~\cite{LCAduan2025quantitative,xu2013mellin,gan2015comparison,MEderbenev2016electrostatic} to derive a $\kappa$-dependent cancellation condition. Recent exact screening-ranged expansions for many dielectric spheres with fixed charge distributions provide a particularly relevant route for extending this analysis beyond an isolated pair~\cite{siryk2025exact}. Modified Poisson--Boltzmann models provide possible continuum descriptions beyond linear screening~\cite{Bbudkov2022modified,Bbudkov2024surface}, whereas strongly coupled or multivalent ions and other solution-mediated effects may require explicit-ion or beyond-mean-field treatments~\cite{Naturewang2024charge,LCAallahyarov1998attraction,LCAwills2024anti,dos2019like,Boriszhang2005long,messina2000strong}. Ionizable surfaces additionally require charge-regulation boundary conditions~\cite{yuan2022impact,Lazlopez2025charge}. Application of the design rule to electrolyte suspensions will therefore require a new cancellation condition that treats ionic screening and dielectric polarization together.

Finally, extending the framework to screened electrostatics and denser assemblies with multiple simultaneous contacts~\cite{Elenaavis2025accommodation,Contactlindgren2024significance} will require identifying when many-body effects limit the accuracy of the pair-derived rule.
Once three-body and higher-order polarization contributions become large enough to perturb the Coulomb-reference structure, quantitative predictions require an explicit many-body treatment~\cite{nitzke2024long,nitzke2026quantitative,lindgren2025charge}.
Future work should establish quantitative diagnostics for when three-body or even higher-order polarization must be included and develop the corresponding corrections for accurate self-assembly predictions.


\begin{acknowledgement}
The authors would like to acknowledge financial support from the Natural Science Foundation of China (Grant No. 12201146) and the Natural Science Foundation of Guangdong Province (Grant No. 2023A1515012197). 
Both authors would like to thank Professor Ho-Kei Chan for insightful discussions.
\end{acknowledgement}

\begin{suppinfo}

\section{Contact energy validation} \label{sec:contact-energy-validation}

This section validates the leading-order contact-energy expression used to generate the analytical design rules in the main text.
The system parameters, calculated contact energies, benchmark reference values, and relative errors are listed in Table~\ref{identical}.

\renewcommand\arraystretch{1.5} 

\begin{table}
	\centering
    \setlength{\tabcolsep}{13.4mm}{
	\begin{tabular}{|c|c|c|c|}
	\hline
	 $k$  & $E_{\text{Ref}}$ & $E_{\text{ele}}$ &  Err$\left(|\frac{E_{\text{ele}}-E_{\text{Ref}}}{E_{\text{Ref}}}|\right)$ \\
	\hline
	 -0.960& 0.563& 0.544& 3.375\% \\
	\hline
	 -0.871& 0.556& 0.541& 2.698\% \\
	\hline
	-0.775& 0.549& 0.537& 2.186\%\\
	\hline
	 -0.678& 0.542& 0.533& 1.661\%\\
	\hline
	 -0.581& 0.535& 0.529& 1.121\% \\
	\hline
	-0.488& 0.529& 0.525& 0.756\%\\
     \hline
     -0.391& 0.523& 0.521& 0.382\% \\
     \hline
     -0.295& 0.516& 0.516& 0 \\
	\hline
	 -0.196& 0.511& 0.511& 0\\
	\hline
	-0.099& 0.505& 0.505 & 0\\
	\hline
	 0& 0.5& 0.5 &0\\
	\hline
	0.042& 0.496& 0.497& 0.202\%\\
    \hline
	 0.188& 0.487& 0.488& 0.205\% \\
	\hline
	0.315& 0.480& 0.480& 0\\
	\hline
 0.476& 0.471& 0.468& 0.637\% \\
	\hline
	0.639& 0.461& 0.454& 1.518\%\\
	\hline
	0.784& 0.452& 0.440& 2.655\%\\
    \hline
    0.960& 0.442& 0.421& 4.751\% \\
    \hline
\end{tabular}}
\caption{Contact-energy validation for a pair of identical dielectric spheres with charges $(Q_1=Q_2=Q)$, dielectric constants $(\epsilon_1=\epsilon_2=\epsilon_\text{in})$, and radii $(a_1=a_2=a)$, suspended in a homogeneous medium with permittivity $\epsilon_\text{out}$. The dielectric contrast is defined as $k=(\epsilon_{\text{in}}-\epsilon_{\text{out}})/(\epsilon_{\text{in}}+\epsilon_{\text{out}})$. Both $E_{\text{Ref}}$ and $E_{\text{ele}}$ are contact energies: $E_{\text{Ref}}$ denotes reference values from the literature~\cite{Bislian2018polarization}, and $E_{\text{ele}}$ denotes the leading-order analytical result used in this work. Both energies are expressed in units of $\frac{Q^2}{4\pi\varepsilon_0\epsilon_{\text{out}}a}$.} 
\label{identical}
\end{table}

Furthermore, we plot the electrostatic force against surface-to-surface distance between two charged polarizable spheres in vacuum in Fig.~\ref{validate}. Both spheres have a radius of $a_1=a_2=1.25$nm, carry central charges of $Q_1 = -1e$ and $Q_2 = -7e$, and dielectric constants $\epsilon_1 = \epsilon_2 = 20$.~\cite{MEderbenev2016electrostatic} Clearly, our theory demonstrates excellent agreement with previous methods. In this setting, we also calculated the contact energy by our formula (297.28) and the Hybrid Method (218.14)~\cite{HMgan2019efficient}, whose energy is on the order of magnitude, indicating that the formula is sufficient to capture the behavior upon contact.

\begin{figure}[htbp]
    \centering
    \includegraphics[scale=1.5]{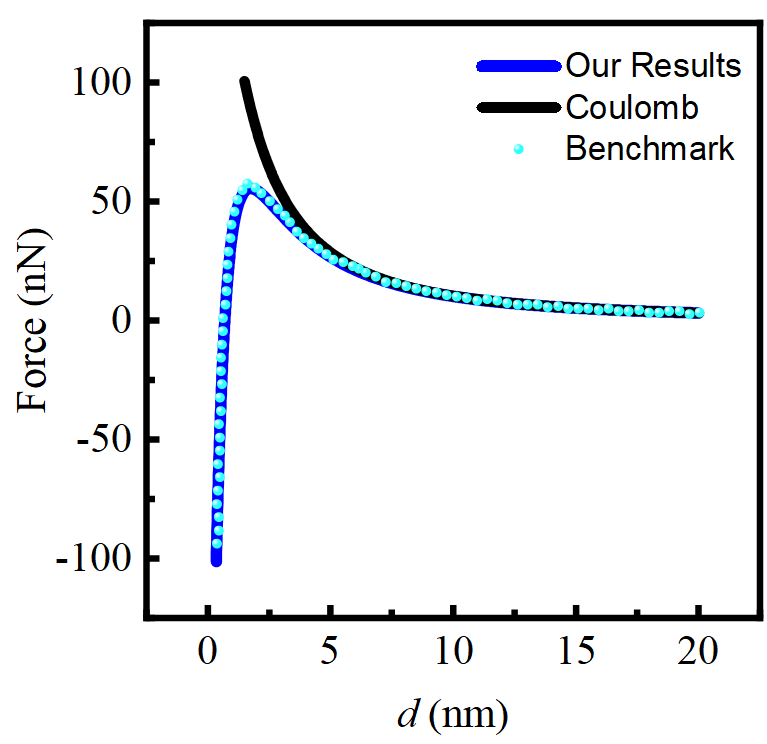}
    \caption{The interaction force between two polarizable spheres in vacuum as a function of sphere-sphere separation $d=R-a_1-a_2$. Both spheres have a radius of $a_1=a_2=1.25 nm$, carry central charges of $Q_1 = -1e$ and $Q_2 = -7e$, and dielectric constants $\epsilon_1 = \epsilon_2 = 20$. Black curve: this work; orange dots: benchmark results from Ref.~\cite{MEderbenev2016electrostatic}; purple curve: the bare Coulomb interaction between the two spheres.}
    \label{validate}
\end{figure}

\section{Analysis for contact energy: the role of asymmetries}

\subsection{Two-sphere systems with charge asymmetry}
We first isolate charge asymmetry by setting the dielectric constants and radii to be symmetric, i.e., $\epsilon_1=\epsilon_2=\epsilon_{\text{in}}$ ($k_1 = k_2 = k$), and $a_1=a_2=a$ ($t_1 = t_2 = t=\frac{a}{2a}=\frac{1}{2}$). Therefore, the contact energy expression in the main text reduces to
\begin{equation}
    E_{\mathrm{ele}} =\frac{Q_1Q_2}{4\pi\varepsilon_0\epsilon_\text{out}
    (2a)} \left\{1-\frac{k}{4}\left(\frac{Q_1}{Q_2}+\frac{Q_2}{Q_1}\right)
     \left[\frac{4}{3} - (1-k) \sum_{n=0}^{\infty}\frac{4^{-n}}{2n+1-k}\right] \right\}\;.
\label{charge}
\end{equation}
Figs.~\ref{LC} and~\ref{OC} illustrate the impact of charge asymmetry on the contact energy of like-charged and oppositely charged spheres, respectively.
In each scenario, we plot the contact energy normalized by the pure Coulomb energy, $E_{\mathrm{ele}}/E_{\mathrm{Coul}}$, versus $(Q_1/Q_2 + Q_2/Q_1)$, with varying dielectric contrast values $k\in(-1, 1)$. The dimensionless parameter $(Q_1/Q_2 + Q_2/Q_1)$, which appears in Eq.~\eqref{charge}, characterizes the strength of charge asymmetry: for like-charged spheres, it reaches its minimum value of 2 when $Q_1 = Q_2$; for oppositely charged spheres, its magnitude increases as the charge asymmetry increases.

Figs.~\ref{LC} and~\ref{OC} show that, in the absence of dielectric contrast ($k=0$), the contact energy is equal to the pure Coulomb energy.
When $k \neq 0$, the contact energy deviates from the Coulomb value as the charge asymmetry or dielectric mismatch increases.
For conductor-like spheres ($k>0$), the normalized contact energy is lower than the Coulomb value, indicating weakened like-charge repulsion or enhanced opposite-charge attraction.
For insulator-like spheres ($k<0$), the normalized contact energy is higher than the Coulomb value, indicating enhanced like-charge repulsion or weakened opposite-charge attraction.
These trends are consistent with earlier studies of polarization-induced like-charge attraction and opposite-charge repulsion~\cite{LCAduan2025quantitative,LCAduan2025mechanisms}.
\begin{figure}[htbp]
    \centering
    \subfigure[]{
        \includegraphics[width=0.47\textwidth]{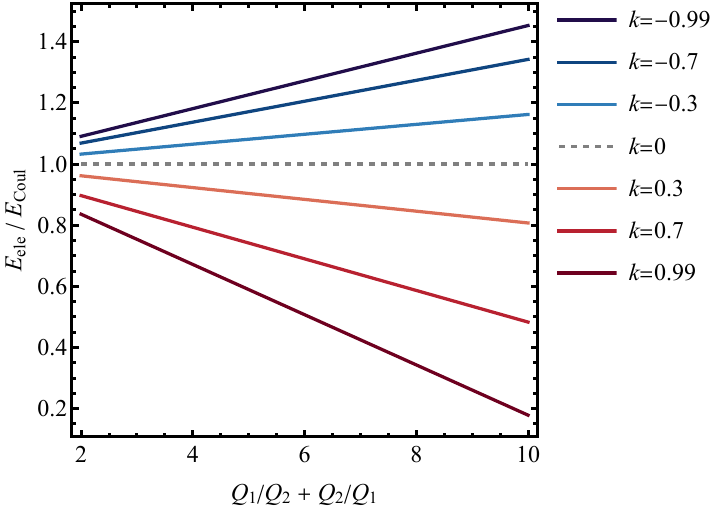}
        \label{LC}
    }
    \subfigure[]{
        \includegraphics[width=0.47\textwidth]{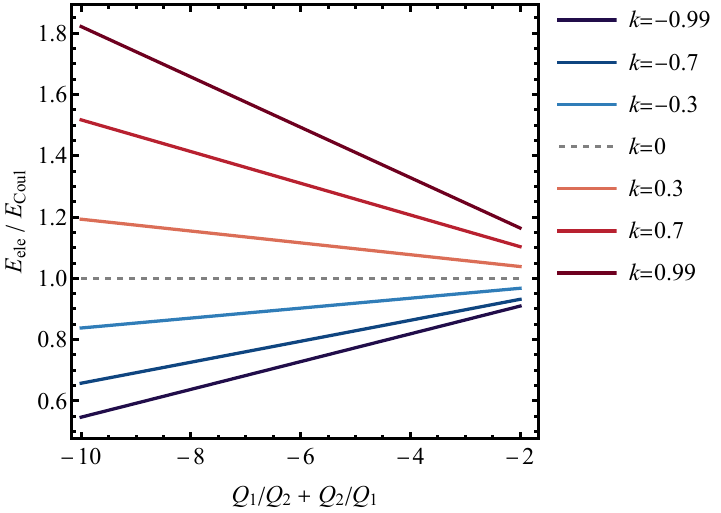}
        \label{OC}
    }
    \caption{Influence of charge asymmetry $\left(\frac{Q_1}{Q_2}+\frac{Q_2}{Q_1}\right)$ on the contact energy of two-sphere systems. (a) Like-charged spheres ($Q_1Q_2>0$) and (b) oppositely charged spheres ($Q_1Q_2<0$). In each scenario, the normalized contact energy $E_{\mathrm{ele}}/E_{\mathrm{Coul}}$ is plotted versus $(Q_1/Q_2 + Q_2/Q_1)$ for different dielectric contrasts $k$.}
    \label{fig:main2}
\end{figure}

\subsection{Two-sphere systems with size asymmetry}
We next isolate size asymmetry by considering a pair of spheres with radii $a_1$ and $a_2$, while setting $Q_1=Q_2=Q$ and $k_1 = k_2 = k$. 
For contacting spheres, the contact ratios satisfy $t_1+t_2=1$, where $t_{1}=a_{1}/(a_1+a_2)$ and $t_{2}=a_{2}/(a_1+a_2)$.
Hence, by substituting $t_2=1-t_1$, the contact energy expression in the main text becomes
\begin{equation}
\begin{split}
    E_{\mathrm{ele}} =\frac{Q^2}{4\pi\varepsilon_0\epsilon_\text{out}
    (a_1+a_2)} \left\{1-\frac{kt_1}{2}
     \left[\frac{1}{1-t_1^2} - (1-k) \sum_{n=0}^{\infty}\frac{t_1^{2n}}{2n+1-k}\right] 
     \right.  \\ \left.      
    -\frac{k(1-t_1)}{2}
     \left[ \frac{1}{t_1(2-t_1)} - (1-k)\sum_{n=0}^{\infty}\frac{(1-t_1)^{2n}}{2n+1-k}\right] \right\}
\end{split}\;.
\label{radius}
\end{equation}

In Fig.~\ref{radii}, we plot the contact energy (normalized by the pure Coulomb energy $E_{\mathrm{Coul}}$) versus $t_1$, with varying dielectric contrast $k\in(-1, 1)$.
Fig.~\ref{radii} shows that, when the two spheres are equal sized $(t_1=t_2 = 0.5)$, the contact energy reaches an extremum for each dielectric contrast $k$.
As the size asymmetry increases ($t_1$ approaches either $0$ or $1$), the contact energy deviates increasingly from the pure Coulomb energy $E_{\mathrm{Coul}}$.
Under extreme dielectric contrast and size asymmetry, the contact energy can become several times larger than the pure Coulomb value or even reverse sign because of strong polarization.
As in Fig.~\ref{fig:main2}, conductor-like spheres $(k > 0)$ weaken the normalized contact energy, whereas insulator-like spheres $(k<0)$ enhance it.
\begin{figure}[htbp]
    \centering
    \includegraphics[scale=1]{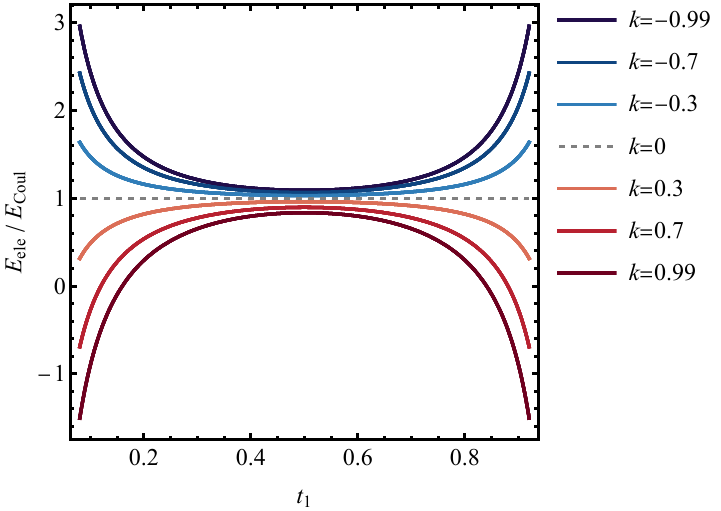}
    \caption{Influence of size asymmetry ($t_1\neq t_2$) on the contact energy of two-sphere systems. The normalized contact energy $E_{\mathrm{ele}}/E_{\mathrm{Coul}}$ is plotted versus $t_1$ for different dielectric contrasts $k\in(-1, 1)$.}
    \label{radii}
\end{figure}

\section{Validation of the pure Coulomb contact force} \label{sec:force-validation}

Tables~\ref{equal-size} and~\ref{unequal-size} document contact-force validation for representative parameter sets predicted by the analytical design rules and marked in the main-text figures.
Table~\ref{equal-size} collects the validation data for two-sphere systems with charge and dielectric asymmetries, whereas Table~\ref{unequal-size} collects the data for systems with size and dielectric asymmetries.

\begin{table}[h!]

	\centering
    \setlength{\tabcolsep}{2.6mm}{
	\begin{tabular}{|c|c|c|c|c|c|c|c|c|}
		\hline
		$\frac{Q_{1}}{Q_{2}}$ & $k_{1}$ & $k_{2}$ & $\epsilon_{1}$ &  $\epsilon_{2}$ & $Q_2$ & $F_{\text{Coul}}$ & $F_{\text{Simul}}$ & Err$\left(\frac{\sqrt{\frac{\sum_{i=1}^{4}\left(F_{\text{Coul}}-F_{\text{Simul}}^{i}\right)^{2}}{4}}}{|F_{\text{Coul}}|}\right)$\\
        
		\hline
		\multicolumn{1}{|c|}{\multirow{4}{*}{0.3}} & 0.065 & -0.932 & 1.139 & 0.040 & 1 & \multicolumn{1}{|c|}{\multirow{4}{*}{0.075}} &  0.073 & \multicolumn{1}{|c|}{\multirow{4}{*}{1.493\%}} \\
		\cline{2-6} \cline{8-8} 
		\multicolumn{1}{|c|}{} &  0.02 & -0.239 & 1.041 & 0.614 & 1 & \multicolumn{1}{|c|}{} & 0.075 & \\
        \cline{2-6} \cline{8-8} 
		\multicolumn{1}{|c|}{} &  -0.065 & 0.587 & 0.878 & 3.843 & 1 & \multicolumn{1}{|c|}{} & 0.074 & \\
		\cline{2-6} \cline{8-8} 
		\multicolumn{1}{|c|}{} &  -0.02 & 0.208 & 0.961 & 1.525 & 1 & \multicolumn{1}{|c|}{} & 0.075 & \\
		\hline
		
		\multicolumn{1}{|c|}{\multirow{4}{*}{0.7}} &  0.09 & -0.199 & 1.198 & 0.668  & 1 & \multicolumn{1}{|c|}{\multirow{4}{*}{0.175}} & 0.174 & \multicolumn{1}{|c|}{\multirow{4}{*}{2.617\%}} \\
		\cline{2-6} \cline{8-8} 
		\multicolumn{1}{|c|}{} & 0.2  & -0.495 & 1.5 & 0.338 & 1 & \multicolumn{1}{|c|}{} & 0.168 & \\
        \cline{2-6} \cline{8-8} 
		\multicolumn{1}{|c|}{} & -0.09  & 0.170 & 0.835 & 1.410 & 1 & \multicolumn{1}{|c|}{} & 0.178 & \\
		\cline{2-6} \cline{8-8} 
		\multicolumn{1}{|c|}{} & -0.2  & 0.347 & 0.667 & 2.063 & 1 & \multicolumn{1}{|c|}{} & 0.170 & \\
		\hline
		
		\multicolumn{1}{|c|}{\multirow{4}{*}{1}} &  0.25 & -0.292 & 1.667  &  0.548 & 1 & \multicolumn{1}{|c|}{\multirow{4}{*}{0.25}} & 0.242 & \multicolumn{1}{|c|}{\multirow{4}{*}{2.020\%}} \\
		\cline{2-6} \cline{8-8} 
		\multicolumn{1}{|c|}{} & 0.1  & -0.106  & 1.222 & 0.808 & 1 & \multicolumn{1}{|c|}{} & 0.249 & \\
        \cline{2-6} \cline{8-8} 
		\multicolumn{1}{|c|}{} & -0.25  & 0.218  & 0.6 & 1.558 & 1 & \multicolumn{1}{|c|}{} & 0.244 & \\
		\cline{2-6} \cline{8-8} 
		\multicolumn{1}{|c|}{} & -0.1  & 0.095  & 0.818 & 1.210 & 1 &  \multicolumn{1}{|c|}{}  & 0.249 & \\
		\hline
		
		\multicolumn{1}{|c|}{\multirow{4}{*}{1.5}} &  0.12  & -0.056 & 1.273 & 0.894 & 1 & \multicolumn{1}{|c|}{\multirow{4}{*}{0.375}} & 0.374 & \multicolumn{1}{|c|}{\multirow{4}{*}{1.347\%}} \\
		\cline{2-6} \cline{8-8} 
		\multicolumn{1}{|c|}{} & 0.3  & -0.153  & 1.857 & 0.735 & 1 & \multicolumn{1}{|c|}{} & 0.367 & \\
        \cline{2-6} \cline{8-8} 
		\multicolumn{1}{|c|}{} & -0.12  & 0.051  & 0.786 & 1.107 & 1 & \multicolumn{1}{|c|}{} & 0.374 & \\
		\cline{2-6} \cline{8-8} 
		\multicolumn{1}{|c|}{} & -0.3  & 0.119 & 0.538 & 1.270  & 1 & \multicolumn{1}{|c|}{} & 0.369 & \\
		\hline
		
		\multicolumn{1}{|c|}{\multirow{4}{*}{2.5}} & 0.45  & -0.085 & 2.626 & 0.843  & 1 & \multicolumn{1}{|c|}{\multirow{4}{*}{0.625}} & 0.614 & \multicolumn{1}{|c|}{\multirow{4}{*}{1.067\%}} \\
		\cline{2-6} \cline{8-8} 
		\multicolumn{1}{|c|}{} &  0.2 & -0.034 & 1.5  & 0.934 & 1 & \multicolumn{1}{|c|}{} & 0.623 & \\
        \cline{2-6} \cline{8-8} 
		\multicolumn{1}{|c|}{} &  -0.45 & 0.063 & 0.379  & 1.134 & 1 & \multicolumn{1}{|c|}{} & 0.618 & \\
		\cline{2-6} \cline{8-8} 
		\multicolumn{1}{|c|}{}  & -0.2  & 0.030  & 0.667  & 1.062 & 1 &  \multicolumn{1}{|c|}{}& 0.623 & \\
		\hline
	\end{tabular}}
	\caption{Contact-force validation for two equal-sized dielectric spheres in vacuum ($\epsilon_{\text{out}}=1$). In the simulations, $Q_2=1\mu C$ and $a_1=a_2=a=1\mu m$, so the force is reported in units of $\frac{1}{4\pi\varepsilon_0}\frac{[C]^2}{[m]^2}$. $F_{\text{Coul}}$ and $F_{\text{Simul}}$ denote the pure Coulomb force and the force calculated using the Hybrid Method~\cite{HMgan2019efficient}, respectively.}
	\label{equal-size}
    
\end{table}

\begin{table}[h!]

	\centering
    \setlength{\tabcolsep}{1.1mm}{
	\begin{tabular}{|c|c|c|c|c|c|c|c|c|c|c|}
		\hline
		$t_1$ & $t_2$ & $a_1$ & $a_2$& $k_{1}$ & $k_{2}$ & $\epsilon_{1}$ &  $\epsilon_{2}$ & $F_{\text{Coul}}$ & $F_{\text{Simul}}$ & Err$\left(\frac{\sqrt{\frac{\sum_{i=1}^{4}\left(F_{\text{Coul}}-F_{\text{Simul}}^{i}\right)^{2}}{4}}}{|F_{\text{Coul}}|}\right)$\\
        
		\hline
		\multicolumn{1}{|c|}{\multirow{4}{*}{0.38}} & \multicolumn{1}{|c|}{\multirow{4}{*}{0.62}} & \multicolumn{1}{|c|}{\multirow{4}{*}{0.613}} & \multicolumn{1}{|c|}{\multirow{4}{*}{1}} & 0.4 & -0.073 & 2.333 & 0.864 & \multicolumn{1}{|c|}{\multirow{4}{*}{0.384}} & 0.383 & \multicolumn{1}{|c|}{\multirow{4}{*}{0.813\%}} \\
		\cline{5-8} \cline{10-10} 
		\multicolumn{1}{|c|}{} & \multicolumn{1}{|c|}{} &  \multicolumn{1}{|c|}{} &  \multicolumn{1}{|c|}{} &  0.15 & -0.025 & 1.353 & 0.951 &  \multicolumn{1}{|c|}{} & 0.385 & \\
        \cline{5-8} \cline{10-10} 
		\multicolumn{1}{|c|}{} & \multicolumn{1}{|c|}{}&  \multicolumn{1}{|c|}{} &  \multicolumn{1}{|c|}{} & -0.4 & 0.055 & 0.429 & 1.116 &  \multicolumn{1}{|c|}{} & 0.378 & \\
		\cline{5-8} \cline{10-10} 
		\multicolumn{1}{|c|}{} & \multicolumn{1}{|c|}{} &  \multicolumn{1}{|c|}{} &  \multicolumn{1}{|c|}{} & -0.15 & 0.022 & 0.739 & 1.045 & \multicolumn{1}{|c|}{} & 0.383 & \\
		
        \hline
        
        \multicolumn{1}{|c|}{\multirow{4}{*}{0.45}} & \multicolumn{1}{|c|}{\multirow{4}{*}{0.55}} & \multicolumn{1}{|c|}{\multirow{4}{*}{0.818}} & \multicolumn{1}{|c|}{\multirow{4}{*}{1}} & 0.25 & -0.131 & 1.667 & 0.768 & \multicolumn{1}{|c|}{\multirow{4}{*}{0.303}} &  0.299 & \multicolumn{1}{|c|}{\multirow{4}{*}{1.082\%}} \\
		\cline{5-8} \cline{10-10} 
		\multicolumn{1}{|c|}{} & \multicolumn{1}{|c|}{} &  \multicolumn{1}{|c|}{} &  \multicolumn{1}{|c|}{} &  0.1 & -0.049 & 1.222 & 0.907 & \multicolumn{1}{|c|}{} & 0.302 & \\
        \cline{5-8} \cline{10-10} 
		\multicolumn{1}{|c|}{} & \multicolumn{1}{|c|}{}&  \multicolumn{1}{|c|}{} &  \multicolumn{1}{|c|}{} & -0.25 & 0.106 & 0.6 & 1.237 & \multicolumn{1}{|c|}{} & 0.298 & \\
		\cline{5-8} \cline{10-10} 
		\multicolumn{1}{|c|}{} & \multicolumn{1}{|c|}{} &  \multicolumn{1}{|c|}{} &  \multicolumn{1}{|c|}{} & -0.1 & 0.045 & 0.818 & 1.094 & \multicolumn{1}{|c|}{} & 0.302 & \\
		\hline

        \multicolumn{1}{|c|}{\multirow{4}{*}{0.5}} & \multicolumn{1}{|c|}{\multirow{4}{*}{0.5}} & \multicolumn{1}{|c|}{\multirow{4}{*}{1}} & \multicolumn{1}{|c|}{\multirow{4}{*}{1}} & 0.15 & -0.164 & 1.353 & 0.718 & \multicolumn{1}{|c|}{\multirow{4}{*}{0.25}} &  0.247 & \multicolumn{1}{|c|}{\multirow{4}{*}{0.775\%}} \\
		\cline{5-8} \cline{10-10} 
		\multicolumn{1}{|c|}{} & \multicolumn{1}{|c|}{} &  \multicolumn{1}{|c|}{} &  \multicolumn{1}{|c|}{} &  0.08 & -0.084 & 1.174 & 0.845 & \multicolumn{1}{|c|}{} & 0.249 & \\
        \cline{5-8} \cline{10-10} 
		\multicolumn{1}{|c|}{} & \multicolumn{1}{|c|}{}&  \multicolumn{1}{|c|}{} &  \multicolumn{1}{|c|}{} & -0.15 & 0.138 & 0.739 & 1.320 & \multicolumn{1}{|c|}{} & 0.248 & \\
		\cline{5-8} \cline{10-10} 
		\multicolumn{1}{|c|}{} & \multicolumn{1}{|c|}{} &  \multicolumn{1}{|c|}{} &  \multicolumn{1}{|c|}{} & -0.08 & 0.076 & 0.852 & 1.165 & \multicolumn{1}{|c|}{} & 0.249 & \\
		\hline

        \multicolumn{1}{|c|}{\multirow{4}{*}{0.54}} & \multicolumn{1}{|c|}{\multirow{4}{*}{0.46}} & \multicolumn{1}{|c|}{\multirow{4}{*}{1}} & \multicolumn{1}{|c|}{\multirow{4}{*}{0.852}} & 0.15 & -0.314 & 1.353 & 0.522 & \multicolumn{1}{|c|}{\multirow{4}{*}{0.292}} &  0.285 & \multicolumn{1}{|c|}{\multirow{4}{*}{1.579\%}} \\
		\cline{5-8} \cline{10-10} 
		\multicolumn{1}{|c|}{} & \multicolumn{1}{|c|}{} &  \multicolumn{1}{|c|}{} &  \multicolumn{1}{|c|}{} &  0.1 & -0.200 & 1.222 & 0.667 & \multicolumn{1}{|c|}{} & 0.288 & \\
        \cline{5-8} \cline{10-10} 
		\multicolumn{1}{|c|}{} & \multicolumn{1}{|c|}{}&  \multicolumn{1}{|c|}{} &  \multicolumn{1}{|c|}{} & -0.15 & 0.245 & 0.739 & 1.649 &  \multicolumn{1}{|c|}{} & 0.288 & \\
		\cline{5-8} \cline{10-10} 
		\multicolumn{1}{|c|}{} & \multicolumn{1}{|c|}{} &  \multicolumn{1}{|c|}{} &  \multicolumn{1}{|c|}{} & -0.1 & 0.170 & 0.818 & 1.410 & \multicolumn{1}{|c|}{} & 0.290 & \\
		\hline
        
        \multicolumn{1}{|c|}{\multirow{4}{*}{0.66}} & \multicolumn{1}{|c|}{\multirow{4}{*}{0.34}} & \multicolumn{1}{|c|}{\multirow{4}{*}{1}} & \multicolumn{1}{|c|}{\multirow{4}{*}{0.515}} & 0.05 & -0.782 & 1.105 & 0.122 & \multicolumn{1}{|c|}{\multirow{4}{*}{0.436}} &  0.424 & \multicolumn{1}{|c|}{\multirow{4}{*}{1.455\%}} \\
		\cline{5-8} \cline{10-10} 
		\multicolumn{1}{|c|}{} & \multicolumn{1}{|c|}{} &  \multicolumn{1}{|c|}{} &  \multicolumn{1}{|c|}{} &  0.02 & -0.271 & 1.041 & 0.574 &  \multicolumn{1}{|c|}{} & 0.432 & \\
        \cline{5-8} \cline{10-10} 
		\multicolumn{1}{|c|}{} & \multicolumn{1}{|c|}{}&  \multicolumn{1}{|c|}{} &  \multicolumn{1}{|c|}{} & -0.05 & 0.513 & 0.905 & 3.107 & \multicolumn{1}{|c|}{} & 0.436 & \\
		\cline{5-8} \cline{10-10} 
		\multicolumn{1}{|c|}{} & \multicolumn{1}{|c|}{} &  \multicolumn{1}{|c|}{} &  \multicolumn{1}{|c|}{} & -0.02 & 0.229 & 0.961 & 1.594 &  \multicolumn{1}{|c|}{} & 0.437 & \\
		\hline
		
	\end{tabular}}
	\caption{Contact-force validation for two unequal-sized dielectric spheres in vacuum ($\epsilon_{\text{out}}=1$). In the simulations, $Q_1=Q_2=1 \mu C$, and $a_1$ and $a_2$ are measured in units of $\mu m$, so the force is reported in units of $\frac{1}{4\pi\varepsilon_0}\frac{[C]^2}{[m]^2}$. $F_{\text{Coul}}$ and $F_{\text{Simul}}$ denote the pure Coulomb force and the force calculated using the Hybrid Method~\cite{HMgan2019efficient}, respectively.}
	\label{unequal-size}
    
\end{table}

\section{Completeness and reproducibility of MD simulations} 

\begin{figure}[htbp]
    \centering
    \subfigure[]{
        \includegraphics[width=0.98\textwidth]{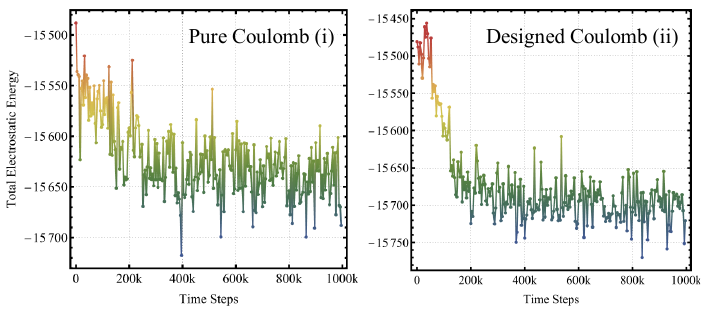}
        \label{1}
    }
    \subfigure[]{
        \includegraphics[width=0.98\textwidth]{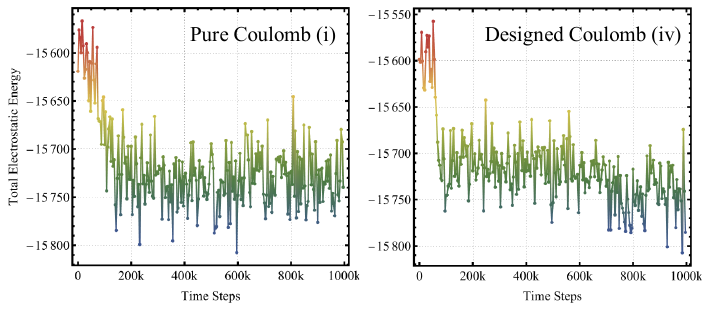}
        \label{2}
    }
    \caption{ Completeness and reproducibility of MD simulations of oppositely charged spheres ($N=200$). (a) Equal-sized systems, where Pure Coulomb (i) and Designed Coulomb (ii) correspond to Fig.5(a) in manuscript. (b) Unequal-sized systems, where Pure Coulomb (i) and Designed Coulomb (iv) correspond to Fig.5(b) in manuscript.
    }
    \label{fig3}
\end{figure}

\end{suppinfo}

\providecommand{\latin}[1]{#1}
\makeatletter
\providecommand{\doi}
  {\begingroup\let\do\@makeother\dospecials
  \catcode`\{=1 \catcode`\}=2 \doi@aux}
\providecommand{\doi@aux}[1]{\endgroup\texttt{#1}}
\makeatother
\providecommand*\mcitethebibliography{\thebibliography}
\csname @ifundefined\endcsname{endmcitethebibliography}
  {\let\endmcitethebibliography\endthebibliography}{}



\end{document}